\documentstyle[aps,prc,epsfig]{revtex}

\def\ee{\end{equation}}
\def\be{\begin{equation}}
\def\eea{\end{eqnarray}}
\def\bea{\begin{eqnarray}}
\def\eeas{\end{eqnarray*}}
\def\beas{\begin{eqnarray*}}
 
\begin{document}
\draft
\title{The pairing Hamiltonian for one pair of identical nucleons bound
in a potential well }
\author{M.B.Barbaro$^{(1)}$, L.Fortunato$^{(2)}$, A.Molinari$^{(1)}$, 
M.R.Quaglia$^{(1)}$}

\address{(1) Dipartimento di Fisica Teorica dell'Universit\`a di Torino, 
via P.Giuria 1, I--10125, Torino, Italy \\
and INFN Sezione di Torino\\
(2) Dipartimento di Fisica ``Galilei'' dell'Universit\`a di Padova, 
via Marzolo 8, I-35131, Padova, Italy\\
and INFN Sezione di Padova}

\maketitle
\begin{abstract}
The problem of one pair of identical nucleons sitting in ${\cal N}$
single particle levels of a potential 
well and interacting through the pairing force 
is treated introducing even Grassmann variables. 
The eigenvectors are analytically 
expressed solely in terms of these with coefficients fixed by the eigenvalues 
and the single particle energies.
When the latter are those of an harmonic oscillator well 
an accurate expression is derived
for both the collective eigenvalue and for those 
trapped in between the single particle levels,
for any strength of the pairing interaction and any number of levels. 
Notably the trapped solutions are labelled through an index upon which they 
depend parabolically.
\end{abstract}


We have recently obtained, in the framework of the Grassmann algebra, 
the analytic expressions of the eigenvalues and the eigenvectors 
of a system of $n$ pairs of like-nucleons interacting through the pairing 
Hamiltonian and sitting in one single-particle level \cite{Bar00}.

We extend the analysis to the case of ${\cal N}$ single-particle levels,
with  energies $e_1, e_2, \cdots e_{\cal N}$
and angular momenta $j_1, j_2, \cdots j_{\cal N}$ 
(all the $j$'s being assumed to be different): here the Hamiltonian, for 
identical particles, reads

\be
H= \sum_{\nu=1}^{\cal N} e_\nu \sum_{m_\nu=-j_\nu}^{j_\nu} 
\lambda^*_{j_\nu m_\nu} \lambda_{j_\nu m_\nu} - 
G \sum_{\mu,\nu=1}^{{\cal N}} 
\sum_{m_\mu=1/2}^{j_\mu}\sum_{m_\nu=1/2}^{j_\nu}
 \varphi^*_{j_\mu m_\mu}  \varphi_{j_\nu  m_\nu}
\label{hamiltonian} \ ,
\ee
where $\lambda_{jm}$ and $\lambda^*_{jm}$ are the odd (anticommuting, 
nilpotent) and\cite{Bar97}

\be
\varphi_{j m} \equiv (-1)^{j-m} \lambda_{j -m} \lambda_{j m} \ ,
\label{phi}
\ee
the even (commuting, nilpotent) Grassmann variables.
The latter is associated with a
pair of fermions with vanishing third component of the total angular momentum
($M$=0).
To start with we confine ourselves to consider one pair only.

Notwithstanding the presence of  
both the $\lambda$'s and the $\varphi$'s, the Hamiltonian
(\ref{hamiltonian}) is diagonalized in the $2{\cal N}$ dimensional basis
\footnote{If a single particle level with $j_\nu=1/2$ ($\Omega_\nu=1$) 
is present, the dimension of the basis is actually $2 {\cal N} -1$,
because, obviously, $\Phi^{(0)}_\nu$ is identically zero.} 

\be
\left\{
\begin{array}{cclc}
\Phi^{(0)}_\nu & = & \frac{1}{\sqrt{\Omega_\nu-1}}
\sum_{m_\nu=1/2}^{j_\nu-1}\varphi_{j_\nu,m_\nu}
\\
\Phi^{(1)}_\nu &=& \varphi_{j_\nu,j_\nu} 
\end{array}
\right. 
\ , \ \ \ \ \ \ \ \ \nu=1,\cdots {\cal N}
\label{Phi}
\ee
which extends the one we have introduced in 
ref.~\cite{Bar00} and describes the two nucleons in the same single-particle
level.
In (\ref{Phi}) $2\Omega_\nu=2 j_\nu+1$ is the degeneracy of the level $j_\nu$.

For one pair, only states with seniority $v$=0 and 2 are allowed.
In the basis (\ref{Phi}) the eigenvalues of the $v$=2 states are trivial, 
whereas for those of the $v$=0 states one recovers
the well-known secular equation

\be
\frac{1}{G}+f(E)=0\ \ \ 
\mbox{with}\ \ \ 
f(E)=\sum_{\nu=1}^{\cal N} 
\frac{\Omega_\nu}{E-2 e_\nu}  \ .
\label{v0}
\ee
The corresponding eigenvectors are

\bea
\psi^{(\nu)}_{v=2}(\Phi^*)&=&\sqrt{1-\frac{1}{\Omega_\nu}}\left\{
\left[\Phi_\nu^{(1)}\right]^*-\frac{1}{\sqrt{\Omega_\nu-1}} 
\left[\Phi_\nu^{(0)}\right]^*\right\}
\label{psiv2}
\eea
and
\bea
\psi^{(\nu)}_{v=0}(\Phi^*) &=& 
\sum_{\mu=1}^{\cal N} w_\mu^{(\nu)} \sqrt{\Omega_\mu-1}
\left\{
\left[\Phi_\mu^{(0)}\right]^*+\frac{1}{\sqrt{\Omega_\mu-1}} 
\left[\Phi_\mu^{(1)}\right]^*\right\} \ ,
\eea
with $\nu=1,\cdots {\cal N}$,
the coefficients $w_\mu^{(\nu)}$ fulfilling the system of equations

\be
\left( E^{(\nu)} - 2 e_\mu \right) w_\mu^{(\nu)} + 
G\sum_{\sigma=1}^{\cal N} \Omega_\sigma w_\sigma^{(\nu)} = 0 \ .
\label{system}
\ee
The above system is easily solved and yields the noticeable formula
(see also \cite{Ric64})
\be
w_{\mu}^{(\nu)}= \frac{E^{(\nu)}-2 e_{\cal N}}{E^{(\nu)}-2 e_\mu}
w_{\cal N}^{(\nu)}\ ,
\label{w}
\ee
which shows that for a given set of single-particle energies the $v$=0 
eigenvectors are fixed by the corresponding eigenvalues.
When $G$ is large (\ref{v0}) develops a collective solution
with large $E$ :
hence the associated $w_\mu$ tend to become all equal and
correspond to a coherent superposition of the so-called $s$-quasibosons.
On the other hand, in the limit $G \rightarrow 0$, where
$E\simeq 2 e_{\nu}$, only one component of the basis, 
i.e. the $\nu$-th one, survives in the wavefunction of the ``trapped'' 
states. 

In general the eigenvalues (and hence the eigenvectors) of (\ref{hamiltonian})
stem from an interplay between the single-particle energies and degeneracies.
Of course this interplay can be numerically explored. Here we pursue the scope 
analytically, when the $e_\nu$ and the $\Omega_\nu$ are available.
One of the few cases where this occurs is for
the harmonic oscillator well, where

\be
e_N = \left(N+\frac{3}{2}\right)\hbar\omega \ \ \ \mbox{and}\ \ \ 
\Omega_N=\frac{1}{2}(N+1)(N+2)\ ,\ \ \ \ \ \ 
N=0,\cdots,\infty\ .
\label{eho}\  
\ee
Accordingly the secular equation (\ref{v0}), 
for ${\cal N}$ levels, becomes

\be
\sum_{N=0}^{{\cal N}-1}\frac{(N+1)(N+2)}{2 N+3-\widetilde{E}}=
\frac{1}{\widetilde{G}}\ 
\label{eqho}
\ee
where $\widetilde{G}=G/2\hbar\omega $
and the energies are measured in units of $\hbar \omega$ 
$(2 {\widetilde{e}}_N=2N+3)$.

\begin{figure}[ht]
\centerline{
\psfig{figure=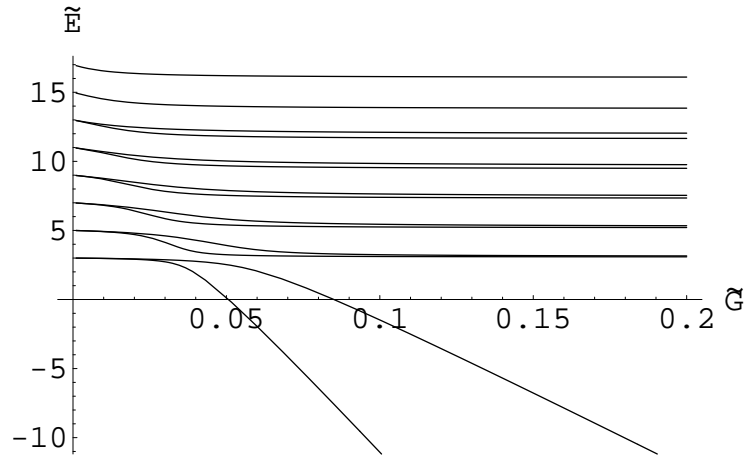,width=10cm,clip=}}
\caption[ ]{The figure shows the solutions $\widetilde E$ of Eq.(\ref{eqho}),
for ${\cal N}$=6 (upper curves) and 8 (lower curves), 
as functions of $\widetilde{G}$.
One can see that with the harmonic oscillator well each trapped solution 
$\widetilde{E}_N$ for $\widetilde{G}>0.1$ tends approximatively to the single 
particle energy $2\widetilde{e}_N$.}
\end{figure}

In Fig.1 the numerical solutions of (\ref{eqho}) are displayed 
for ${\cal N}$=6 and 8 versus $\widetilde{G}$. 
Remarkably, the dependence upon $\widetilde{G}$ is seen to be lost 
for $\widetilde{G}\ge 0.1$. Furthermore in this regime the trapped
solutions are mildly depending upon by ${\cal N}$.

Although aware that the solutions of (\ref{eqho}) cannot, in general, be 
algebraically expressed (for ${\cal N}\geq$ 5), we explore whether
the simple ansatz
\be
\widetilde{E}_{\bar N}=a {\bar N}^2+b {\bar N}+c 
\ \ \ \ \ \ \ \ \ \ \ \ \ \ {\bar N}=0,\cdots {\cal N}-2 
\label{parab}
\ee
provides a good representation of the trapped solutions
(the collective solution $\widetilde{E}_{c}$ will be separately treated).

To fix the coefficients $a$, $b$ and $c$, we recast (\ref{eqho}) in the 
polynomial form 

\be
\widetilde{E}^{\cal N}+a_1 \widetilde{E}^{{\cal N}-1}+
a_2 \widetilde{E}^{{\cal N}-2}+
\cdots a_{{\cal N}-1} \widetilde{E}+a_{\cal N}=0 \ ,
\label{eqk}
\ee
finding for the first three coefficients the expressions

\bea
a_1 &=& \frac{1}{3} {\cal N}({\cal N}+2) [\widetilde G({\cal N}+1)-3]
\\
a_2 &=&  \frac{1}{6} {\cal N}({\cal N}-1) 
[-\widetilde G ({\cal N}+1)({\cal N}+2)(2{\cal N}+3)+
3{\cal N}^2+11 {\cal N}+11]
\\
a_3 &=&  \frac{1}{90} {\cal N}({\cal N}-1)({\cal N}^2-4) 
[\widetilde G({\cal N}+1)(15 {\cal N}^2+40{\cal N}+27)-
15({\cal N}^2+3 {\cal N}+3)]\ .
\nonumber
\\
\eea

Then the first three Viete equations, namely

\bea
\sum_{\bar N=0}^{{\cal N}-2} \widetilde{E}_{\bar N} &=& -a_1-\widetilde{E}_c
\label{Vieta1}
\\
\sum_{\bar N=0}^{{\cal N}-2} \widetilde{E}^2_{\bar N} &=& 
a_1^2-2 a_2-\widetilde{E}_c^2
\label{Vieta2}
\\
\sum_{\bar N=0}^{{\cal N}-2} \widetilde{E}^3_{\bar N} &=& 
-3 a_3-a_1(a_1^2-3 a_2)- 
\widetilde{E}_c^3 \ ,
\label{Vieta3}
\eea
yield a non-linear system in the unknowns $a$, $b$ and $c$, if
$\widetilde{E}_c$ is known. 
This system can be solved by expressing, via eq.(\ref{Vieta1}), 
$c$ as a function of $a$ and $b$ 
\bea
c(a,b)&=&-\frac{1}{{\cal N}-1}\left\{\widetilde{E}_c+\frac{b}{2} ({\cal N}-1)
({\cal N}-2)+\frac{a}{6} ({\cal N}-1)({\cal N}-2)(2{\cal N}-3) \right.
\nonumber\\
&+&
\left.\frac{1}{3} {\cal N}({\cal N}+2)
\left[{\widetilde G}({\cal N}+1)-3\right]\right\}\ .
\label{eqc}
\eea
In turn (\ref{eqc}), inserted into (\ref{Vieta2}), yields $b$ as 
a function of $a$. One finds
\be
b(a) = -\frac{ 15 a({\cal N}^4 -6 {\cal N}^3+13 {\cal N}^2-12 {\cal N}+4)
+\sqrt{\Delta}}{15 ({\cal N}-1)^2 ({\cal N}-2)}
\ee
with
\bea
\Delta &=&  -15 ({\cal N}-1)^2 ({\cal N}-2) \left\{ a^2 ({\cal N}-1)^2
({\cal N}-1)({\cal N}-2)({\cal N}-3)
\right.\\
 &+& 20 \left[9 \widetilde{E}_c^2 + 6 \widetilde{E}_c
({\cal N}+2)(\widetilde{G}{\cal N}+\widetilde{G}-3)-3({\cal N}^3-4 {\cal N}^2
-13 {\cal N}-11)
\right.\nonumber\\
 &-& \left. \left. \widetilde{G}^2 {\cal N}({\cal N}-2)({\cal N}+1)^2
({\cal N}+2)^2 + 3 \widetilde{G}({\cal N}+1)({\cal N}+2)({\cal N}^2-
4 {\cal N}-3)\right]\right\}\ .
\nonumber
\eea

Finally, from (\ref{Vieta3}), a cumbersome equation for $a$ follows, 
not reported here.
While (\ref{Vieta1}) and (\ref{Vieta2})  are easily solved analytically,
the non linear equation
(\ref{Vieta3}) for $a$ can only be solved numerically and admits, 
in general, several solutions. 
The one appropriate for our problem is
selected by requiring that the trapped energies lye in between the 
single-particle levels of the harmonic oscillators.
Moreover,
it should be pointed out that, owing to the high degree of non-linearity of 
(\ref{Vieta3}), this solution 
turns out to be extremely sensitive to the collective energy $\widetilde{E}_c$,
especially  when $\widetilde G$ is large.

Hence an accurate expression for the collective energy is needed.
We look for the latter in the domain of small $\widetilde G$
(say $\widetilde G$=0.05 - 0.1), as it follows from the empirical
determination of $G$ in atomic nuclei and from the experimental 
nuclear single-particle levels\cite{Fesh}.

For this scope we start by realizing that
${\widetilde E}_c(0, {\cal N})$=3 and
${\widetilde E}_c(\widetilde G_0, {\cal N})$=0, being

\be
\widetilde{G}_0=\left[\sum_{N=0}^{{\cal N}-1}\frac{(N+1)(N+2)}
{\left(2N+3\right)}\right]^{-1}\ . 
\ee

Moreover

\be
\left.\frac{\partial \widetilde E_c(\widetilde G,{\cal N})}
{\partial \widetilde G}
\right |_{{\widetilde G}=0} = -2
\ee     
and
\be
\left.\frac{\partial \widetilde E_c(\widetilde G,{\cal N})}
{\partial \widetilde G}
\right |_{{\widetilde G}={\widetilde G}_0} =
-\left[\sum_{N=0}^{{\cal N}-1}\frac{(N+1)(N+2)}{2N+3}\right]^2
\left[\sum_{N=0}^{{\cal N}-1}\frac{(N+1)(N+2)}{
{\left(2N+3\right)}^2}\right]^{-1} \ .
\ee     

The above constraints are fulfilled by the cubic function (in $\widetilde G$)
\be
\widetilde E_c^{(0)}(\widetilde G,{\cal N})= 3-2\widetilde G -
\left[9+\left(
\left.\frac{\partial \widetilde E_c}{\partial \widetilde G}
\right |_{ {\widetilde G}_0}
-4\right)\widetilde G_0\right]
\frac{\widetilde G^2}{\widetilde G_0^2} + 
\left[6+\left(
\left.\frac{\partial \widetilde E_c}{\partial \widetilde G}
\right |_{{\widetilde G}_0}
-2\right)\widetilde G_0 \right]
\frac{\widetilde G^3}{\widetilde G_0^3}\ , 
\label{Ecubic}
\ee
which thus provides an excellent representation of the 
collective energy (see Table 1). 
If an even better $\widetilde E_c$ is wished, one can proceed perturbatively
setting

\be
\widetilde{E}_c = \widetilde{E}_c^{(0)} + \delta 
\ee
and expanding in the very small parameter $\delta/M(N)$ where
\be
M(N)= 2N+3 -\widetilde{E}_c^{(0)}\ .
\ee
One thus gets the recursive relation

\be
\widetilde{E}_c^{(k+1)}=\widetilde{E}_c^{(k)} +
\left[\frac{1}{\widetilde G}- 
\sum_{N=0}^{{\cal N}-1}\frac{(N+1)(N+2)}{2N+3-\widetilde{E}_c^{(k)}}\right]
\left[\sum_{N=0}^{{\cal N}-1}\frac{(N+1)(N+2)}{
{\left(2N+3-\widetilde{E}_c^{(k)}\right)}^2}
\right]^{-1}\ .
\label{Erecursion}
\ee  
The energies provided by (\ref{Erecursion}) fastly converge to the exact 
solution, as shown 
in Table~1, but, as above mentioned, a high precision is required,
which obtains after 4 iterations.

\begin{table}[h]
\begin{center}
\begin{tabular}{c|c|c|c|c|c|c}
& \multicolumn{6}{c}{$\widetilde G=0.05$}\\ \hline
${\cal N}$ & $\widetilde{E}_c^{(0)}$  &  $\widetilde{E}_c^{(1)}$  
& $\widetilde{E}_c^{(2)}$  & $\widetilde{E}_c^{(3)}$ & $\widetilde{E}_c^{(4)}$
& $\widetilde{E}_c^{(e)}$       \\
\hline
2    & 2.87617 & 2.88394 & 2.88349 & 2.88348 & 2.88348  & 2.88348  \\ 
3    & 2.81180 & 2.87628 & 2.86196 & 2.86014 & 2.86012  & 2.86012  \\
4    & 2.70198 & 2.89185 & 2.84845 & 2.82471 & 2.82056  & 2.82046  \\
5    & 2.53580 & 2.86975 & 2.80333 & 2.75484 & 2.74104  & 2.74028  \\
6    & 2.24594 & 2.63571 & 2.54951 & 2.52958 & 2.52886  & 2.52886  \\
7    & 1.61587 & 1.84837 & 1.83108 & 1.83094 & 1.83094  & 1.83094  \\
8    & .134714 & .136135 & .136135 & .136135 & .136135  & .136135 \\
\hline
& \multicolumn{6}{c}{$\widetilde G=0.1$}\\ \hline
${\cal N}$ & $\widetilde{E}_c^{(0)}$  &  $\widetilde{E}_c^{(1)}$  
& $\widetilde{E}_c^{(2)}$  & $\widetilde{E}_c^{(3)}$ & $\widetilde{E}_c^{(4)}$
& $\widetilde{E}_c^{(e)}$       \\
\hline
2          &  2.70757   &  2.72972 & 2.72822 & 2.72822 & 2.72822 &  2.72822 \\ 
3          &  2.45586   &  2.61626 & 2.58336 & 2.58335 & 2.58335 &  2.58335  \\
4          &  1.96617   &  2.21918 & 2.18238 & 2.18238 & 2.18238 &  2.18238 \\
5          &  0.919942  &  1.00125 & 1.00000 & 1.00000 & 1.00000 &  1.00000 \\
6          & -1.66966   & -1.47625 &-1.48025 &-1.48025 &-1.48025 & -1.48025  \\
7          & -8.37559   & -4.88888 &-5.44566 &-5.44568 &-5.44568 & -5.44568  \\
8          & -24.4931   & -3.26859 &-10.7011 &-11.0491 &-11.0546 & -11.0546  \\
\end{tabular}
\caption{Comparison between the exact $\widetilde E_c^{(e)}$ and 
the approximate $\widetilde E_c^{(k)}$ 
[eq.(\ref{Ecubic}) and eq.(\ref{Erecursion})] collective energies  
for some values of ${\cal N}$ and $\widetilde{G}$=0.05 and 0.1}
\end{center}
\end{table}

Formula (\ref{Erecursion}) yields $\widetilde{E}_c$ also when  
$\widetilde{G}$ is large. Here however the analogous of (\ref{Ecubic})
follows by expanding 
Eq.(\ref{eqho}) in the parameter $(2N+3)/\widetilde{E}$.
One thus gets for collective energy, expanded up to terms $1/{\widetilde{G}}$,
the expression

\be
\widetilde{E}_c^{(0)} = -\frac{\widetilde{G}}{3} {\cal N}({\cal N}+1)
({\cal N}+2) +\frac{3}{2} ({\cal N}+1) -
\frac{9({\cal N}-1) ({\cal N}+3)}{20{\cal N}({\cal N}+1)({\cal N}+2)}
 \frac{1}{\widetilde{G}} \ .
\label{Ecappr}
\ee
The above, when inserted in (\ref{Erecursion}), 
yields results as accurate as those obtained in the domain of small 
$\widetilde G$.

With the collective energy fixed, the coefficients $a$, $b$ and $c$ 
can be found. 
We quote in Table~2, as an example, our predictions for the eigenvalues of 
the pairing hamiltonian for one pair in the ${\cal N}=5$ case, 
using as input (\ref{Ecappr}) when $\widetilde G=1$ and $\widetilde G=5$ 
and (\ref{Ecubic}) when $\widetilde G=0.05$ and $\widetilde G=0.1$.
Our results are seen to agree with the exact ones 
obtained via the numerical solution of (\ref{eqho}) to better than $0.27\%$.   
This occurs as well for all the cases we have explored.
Thus the simple ansatz (\ref{parab}) is remarkably accurate.
Furthermore the solutions (\ref{parab}), when $\widetilde G$ is large, scale
in $\widetilde G$. Indeed, in this condition, the right-hand sides
of the Vieta equations (\ref{Vieta1}-\ref{Vieta3}) are easily seen  to be
$\widetilde G$-independent when the collective solution is given by 
(\ref{Ecappr}). 

\begin{table}[h]
\begin{center}
\begin{tabular}{c|c|c|c|c|c|c|c|c}
 & \multicolumn{2}{c|}{$\widetilde G=0.05$}& 
\multicolumn{2}{c|}{$\widetilde G=0.1$}
 & \multicolumn{2}{c|}{$\widetilde G=1$}& 
\multicolumn{2}{c}{$\widetilde G=5$}\\ \hline
$\bar N$ & $\widetilde{E}_{\bar N}^{(e)}$  &  $\widetilde{E}_{\bar N}^{(app)}$
& $\widetilde{E}_{\bar N}^{(e)}$  &  $\widetilde{E}_{\bar N}^{(app)}$    
& $\widetilde{E}_{\bar N}^{(e)}$  &  $\widetilde{E}_{\bar N}^{(app)}$    &
$\widetilde{E}_{\bar N}^{(e)}$  &  $\widetilde{E}_{\bar N}^{(app)}$    \\
\hline
0           &  4.2872 &  4.2812       
            &  3.4245 &  3.4266
            &  3.1583 &  3.1601 
            &  3.1493 &  3.1510 \\ 
1           &  6.0892 &  6.1056    
            &  5.6302 &  5.6237
            &  5.3673 &  5.3621 
            &  5.3524 &  5.3472 \\
2           &  8.1171 &  8.1021     
            &  7.8422 &  7.8485
            &  7.6136 &  7.6190 
            &  7.5965 &  7.6015\\
3           &  10.266 &  10.271     
            &  10.103 &  10.101
            &  9.9314 &  9.9297 
            &  9.9157 &  9.9140 \\
\end{tabular}
\caption{Comparison between exact $(e)$ ``trapped'' solutions of 
Eq.(\ref{eqho}) and approximate $(app)$ ones, obtained from the ansatz
(\ref{parab}) for ${\cal N}=5$ levels. 
The coefficients $(a,b,c)$ of the parabola are (0.086,1.738,4.281) 
when $\widetilde{G}=0.05$, 
(0.014,2.183,3.427) when $\widetilde{G}=0.1$,
(0.027,2.175,3.160) when $\widetilde{G}=1$ 
and (0.029,2.167,3.151) when $\widetilde{G}=5$.}  
\end{center}
\end{table}

In conclusion, 
although the pairing problem can be, of course, solved numerically, yet
we believe that our semi-analytical solution might be of some help for
treating the situation when $n$ pairs, sitting in an 
harmonic oscillator well, are present.
Also interesting appears the extension of the present analysis to the 
situation where the pair is made out of a neutron and a proton, particularly
when these are in an isospin singlet state \cite{Gar01}. 
In this case indeed the partners, in order to feel the pairing interaction, 
are forced to seat in different shells, at least $2 \hbar \omega$ apart.
Whether in these conditions our semianalytical solution holds valid as well 
and a collective mode eventually develops is an issue we are currently 
exploring.

\end{document}